\documentclass{aa}  
\usepackage{graphicx}
\usepackage{txfonts}
\usepackage{booktabs}
\begin{document}

   \title{The globular cluster VVV~CL002 falling down to the hazardous Galactic centre}


   \author{Dante Minniti \inst{1,2,3}
       \and
           Noriyuki Matsunaga \inst{4,5}
       \and
           Jos\'e G. Fern\'andez-Trincado \inst{6} 
       \and
           Shogo Otsubo \inst{5} 
       \and
           Yuki Sarugaku \inst{5} 
       \and
           Tomomi Takeuchi \inst{5} 
       \and
           Haruki Katoh \inst{5} 
       \and
           Satoshi Hamano \inst{7} 
       \and
           Yuji Ikeda \inst{5,8} 
       \and
           Hideyo Kawakita \inst{5,9} 
       \and
           Philip W. Lucas \inst{10} 
       \and
           Leigh C. Smith \inst{11} 
       \and
           Ilaria Petralia \inst{1} 
       \and
           Elisa Rita Garro \inst{1} 
       \and
           Roberto K. Saito \inst{3} 
       \and
           Javier Alonso-Garc\'ia \inst{12} 
       \and
           Mat\'ias G\'omez \inst{1} 
       \and
           Mar\'ia Gabriela Navarro \inst{13} 
    }
    
   \institute{Instituto de Astrof\'isica, Depto. de Ciencias F\'isicas, Facultad de Ciencias Exactas, Universidad Andres Bello, Fernandez Concha 700, Santiago, RM, Chile  \\
       \email{dante.minniti@unab.cl} 
   \and
       Specola Vaticana, Vatican Observatory, Castelgandolfo, V00120, Stato Citta Vaticano, Italy
   \and
       Departamento de Fisica, Universidade Federal de Santa Catarina, Florianopolis, Trindade 88040-900, SC, Brazil
   \and
       Department of Astronomy, School of Science, The University of Tokyo, 7-3-1 Hongo, Bunkyo-ku, Tokyo 113-0033, Japan \\
        \email{matsunaga@astron.s.u-tokyo.ac.jp}
   \and
       Laboratory of Infrared High-resolution Spectroscopy (LiH), Koyama Astronomical Observatory, Kyoto Sangyo University, Motoyama, Kamigamo, Kita-ku, Kyoto 603-8555, Japan
   \and
       Instituto de Astronom\'ia, Universidad Cat\'olica del Norte, Av. Angamos 0610, Antofagasta, Chile
   \and
       National Astronomical Observatory of Japan, 2-21-1 Osawa, Mitaka, Tokyo 181-8588, Japan
   \and
       Photocoding, 460-102 Iwakura-Nakamachi, Sakyo-ku, Kyoto 606-0025, Japan
   \and
       Department of Astrophysics and Atmospheric Sciences, Faculty of Science, Kyoto Sangyo University, Motoyama, Kamigamo, Kita-ku, Kyoto 603-8555, Japan
   \and
       Centre for Astrophysics Research, University of Hertfordshire, College Lane, Hatfield, AL10 9A,  United Kingdom
   \and
       Institute of Astronomy, University of Cambridge, Madingley Rd., Cambridge, CB3 0HA,  United Kingdom
   \and
       Centro de Astronomia (CITEVA), Universidad de Antofagasta, Av. Angamos 601, Antofagasta,  Chile
   \and
       INAF, Osservatorio Astronomico di Roma, Via di Frascati 33, Monteporzio Catone, 00040, Italy
   }

   \date{Received Month DD, Year; accepted Month DD, Year}

 
  \abstract
   {The Galactic centre is hazardous for stellar clusters because of the strong tidal force. Supposedly, many clusters were destroyed and contributed stars to the crowded stellar field of the bulge and the nuclear stellar cluster.  However, it is hard to develop a realistic model to predict the long-term evolution of the complex inner Galaxy, and observing surviving clusters in the central region would provide crucial insights into destruction processes. }
   {Among hitherto-known Galactic globular clusters, VVV~CL002 is the closest to the centre, 0.4~kpc, but has a very high transverse velocity, 400\,km\,s$^{-1}$. The nature of this cluster and its impact on Galactic astronomy need to be addressed with spectroscopic follow-up.}
   {Here we report the first measurements of its radial velocity and chemical abundance based on near-infrared high-resolution spectroscopy.}
   {We found that this cluster has a counterrotating orbit constrained within 1.0\,kpc of the centre, as close as 0.2\,kpc at the perigalacticon, confirming that the cluster is not a passerby from the halo but a genuine survivor enduring the harsh conditions of the Galactic mill's tidal forces. In addition, its metallicity and $\alpha$ abundance ([$\alpha$/Fe] $\simeq +0.4$ and [Fe/H]$=-0.54$) are similar to some globular clusters in the bulge. Recent studies suggest that stars with such $\alpha$-enhanced stars were more common at 3--6\,kpc from the centre around 10 Gyrs ago.}
   {We infer that VVV~CL002 was formed outside but is currently falling down to the centre, exhibiting a real-time event that must have occurred to many clusters a long time ago.}


   \maketitle
%

\section{Introduction}

Whereas more than 200 globular clusters have been found today in the Galaxy,
there is plenty of evidence for clusters having been destroyed by various evolutionary and dynamical processes, 
including dynamical friction, shocking by disk and bulge,
tidal disruption, and so on
\citep{Leon-2000, Murali-1997, Gnedin-1997, Baumgardt-2003, Moreno-2022}.
Many of these dynamical processes are stronger in the deep potential well of the inner Galaxy.
Numerical simulations have revealed that the supermassive black hole Sagittarius A*
\citep{Ghez-2008, Genzel-2010} 
is a very efficient machine of grinding clusters,
where globular clusters could be rapidly demolished \citep{Arca-Sedda-2017, Navarro-2023}.
In order to address the process of globular cluster disruption,
it is crucial to understand not only clusters that have been destroyed
but also surviving globular clusters. 
Search for globular clusters in the inner Galaxy has been 
incomplete because of large interstellar extinction and heavy source crowding,
Still, recent near-infrared surveys have revealed dozens of candidate clusters
in the inner bulge \citep{MoniBidin-2011,Borissova-2014}.
Then, confirmation of member stars and detailed characterization of
the clusters' properties require infrared spectroscopic observations due to large 
interstellar extinction near the centre.

VVV~CL002 is a relatively low-luminosity globular cluster
($M_K=-7.1$~mag, $M_V=-4.6$~mag \citep{Minniti-2021}), which was
discovered at 1.1~deg from the Galactic centre through  
the VISTA Variables in the Via Lactea (VVV) survey \citep{MoniBidin-2011}.
The distance measured with the red clump places this cluster 
closest to the centre, 0.4~kpc, among known globular clusters,
and an additional surprise is the high transverse velocity, 400\,km\,$^{-1}$,
inferred from the proper motion \citep{Minniti-2021}.
A part of RR~Lyr variables with such high velocities found in 
the bulge turned out to be halo objects \citep{Kunder-2020}.
Is VVV~CL002 an interloper from the halo passing near the Galactic centre?
If it remains near the centre instead, we need to
consider how it is surviving without being tidally disrupted.
Spectroscopic observations are required to answer these questions.

\section{Observation and data analysis}

\subsection{Target selection}

\begin{figure}[htbp]%
\centering
\includegraphics[width=\hsize]{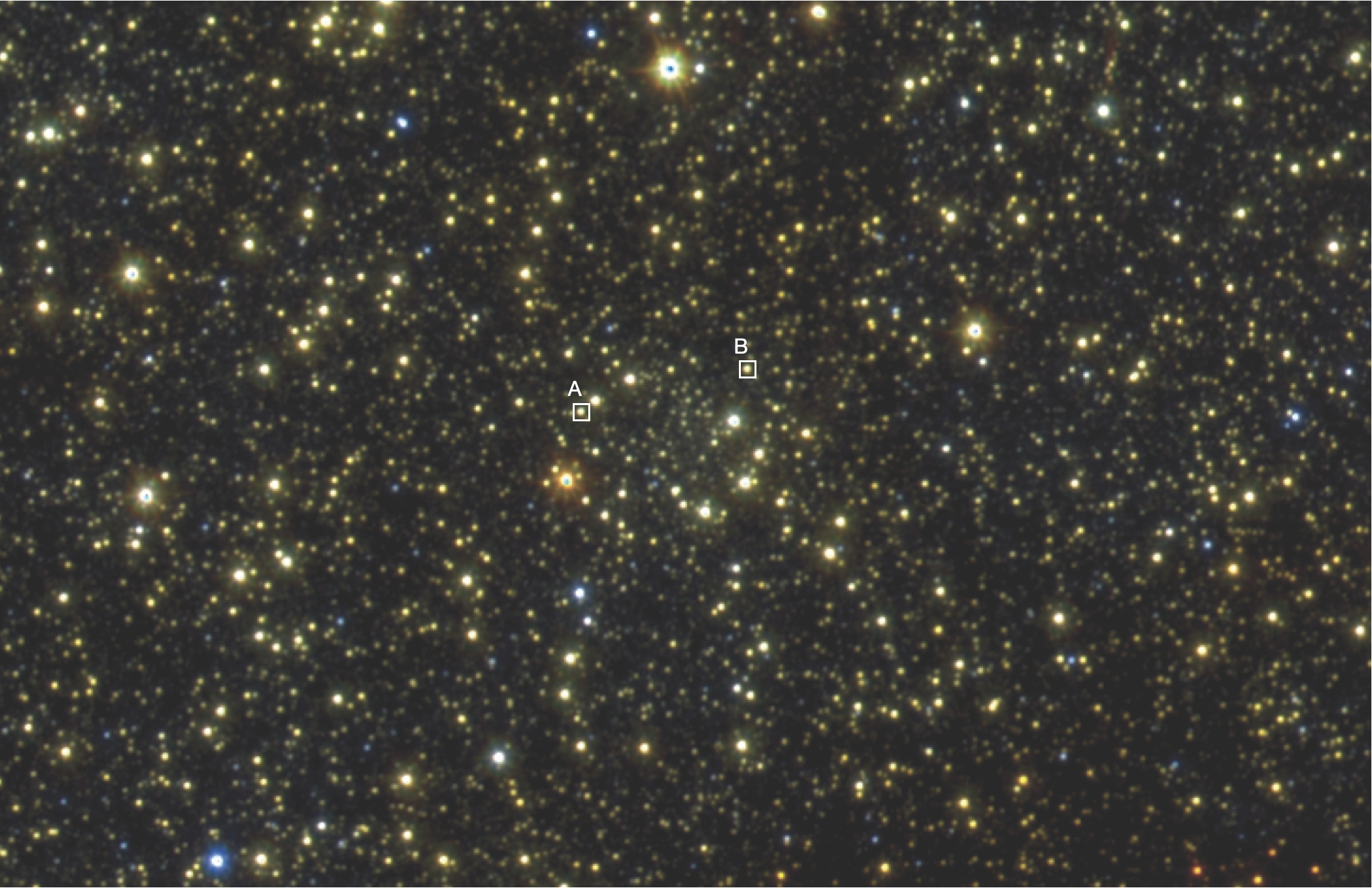}
\caption{
Near-infrared image of the field around VVV~CL002 made with VVV images in the $JHK_\mathrm{s}$-band filters, indicating the target red giant stars A (VIRAC2 ID=9327562027332) and B (VIRAC2 ID=9327562015390). 
The field covers $4'\times 3'$, oriented along Galactic coordinates, with latitude increasing upwards and longitude to the left. 
The globular cluster is located at the centre of this image, that also illustrates the overwhelming density of field stars. 
}\label{fig:finding_chart}
\end{figure}

\begin{figure*}[htbp]%
\centering
\includegraphics[width=\hsize]{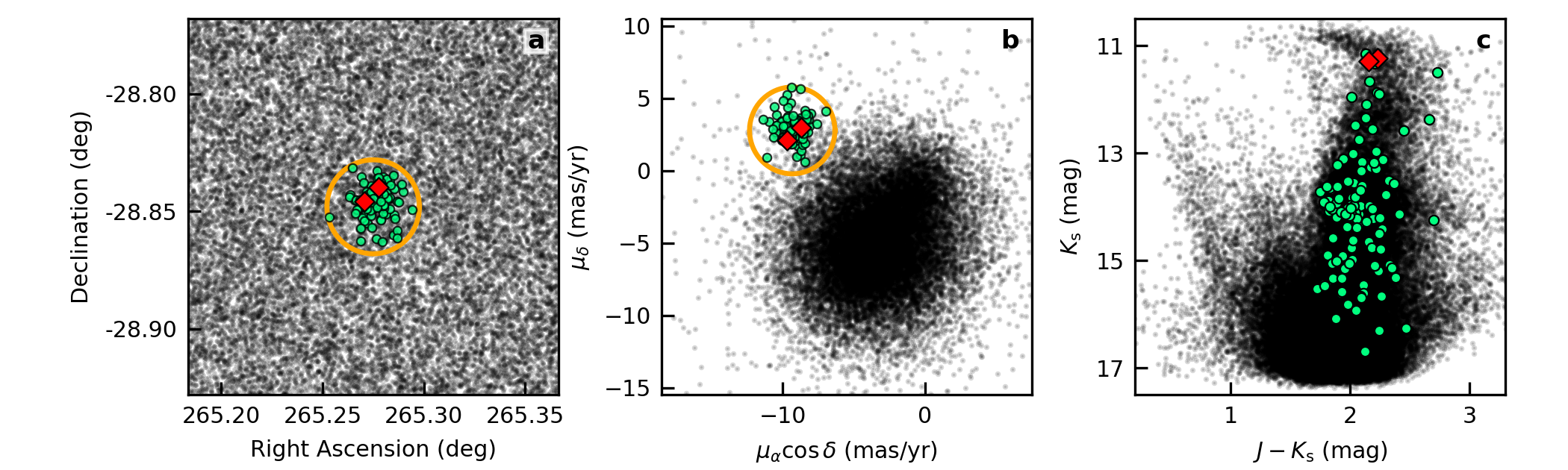}
\caption{Illustration of the target selection from VVV CL002: (a)~the field surrounding the cluster, (b)~the proper motion diagram, and (c) the near-IR color-magnitude diagram. The field stars are plotted with the black points, while stars located inside the selection circles (orange) on both the panels (a) and (b) are indicated by green circles. A part of these stars in green were selected as spectroscopic targets based also on panel (c), and we observed the two red giants indicated by red diamonds.
}\label{fig:virac}
\end{figure*}

Embedded in the crowded stellar field, selecting good targets
for follow-up spectroscopic observations is a crucial step.
In particular, the cluster membership determination is very tricky in such a crowded field (Fig.~\ref{fig:finding_chart}).
The line-of-sight contamination of stars in various groups increase at low latitudes, and accurate 6D information (position, radial velocity and proper motion) is vital in judging membership.
Based on the VVV's photometry and proper motions in the updated VIRAC2 database (Smith et al., in prep) 
we selected a few candidate members of red giant
(Fig.~\ref{fig:virac})
including the two stars that we actually observed.
Therefore we initially selected the four brightest RGB stars with high probability of membership according to the photometric and astrometric data. The main criteria for the selection of this sample were:
1) Stars within 0.1 mag from the mean RGB ridge line in the near-IR color-magnitude diagram.
2) Stars with PMs within 0.1 mas/yr from the mean GC PM.
3) Stars that appeared unblended in the optical and near-IR images.
4) Stars brighter than J = 14.2 mag.
Two out of these four main targets selected were successfully observed (Table 1). The remaining two targets could not be observed due to time and weather constraints during our observing run. These are Gaia ID 9327562004074, with Ks=11.87 mag, J-Ks=2.22 mag; and 
Gaia ID 9327562025638 , with Ks=11.90 mag, J-Ks=2.24 mag, which would also be prime targets for future spectroscopic observations.
The proper motions of candidate members of VVV~CL002
show a large offset from the bulk of bulge field stars,
which indicates the large transverse velocity 
of the clusters, {$\sim$}400\,km\,s$^{-1}$ \citep{Minniti-2021}.
Although some contaminants remain,
stars selected by the proper motion show the
red giant branch and red clump expected for
a globular cluster.


  \begin{figure*}
   \centering
   \includegraphics[width=0.7\hsize]{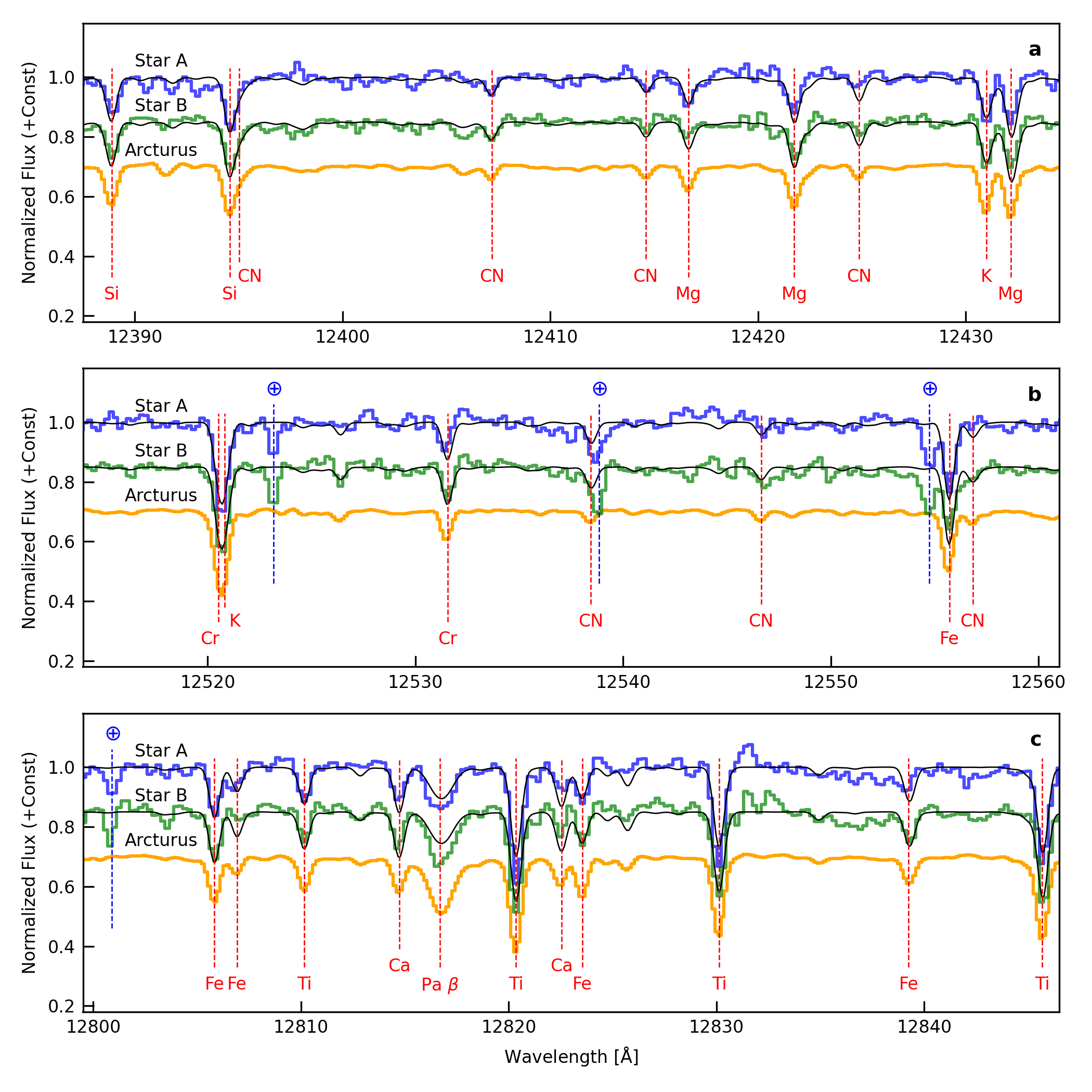}
   \caption{
   Spectra for the two red giant members of the globular cluster VVV~CL002: star A in blue and star B in green. Three representative parts containing absorption lines used for abundance measurements (Mg, Si, and Fe) and other lines are presented together with model spectra synthesized with $T_\mathrm{eff}=4100$\,K, $\log g=1.5$ and [Fe/H]$=-0.54$ (black). 
   The spectrum of Arcturus (orange) is also shown for comparison,
   shifted to the appropriate velocity to match with the spectra of target stars. 
   Arcturus ($T_\mathrm{eff}=4280$~K, [Mg/H]$=-0.2$, [Si/H]$=-0.2$, [Fe/H]$=-0.5$) is a prototype red giant
   observed with the same setup \citep{Fukue-2021}.
   Note that all the spectra look very similar to each other.
   Absorption lines with the $\oplus$ mark show telluric lines, which are only seen in the two target stars but not in Arcturus for which the telluric correction was done.
   }\label{fig:spectra}
   \end{figure*}


\begin{figure}[htbp]%
\centering
\includegraphics[width=\hsize]{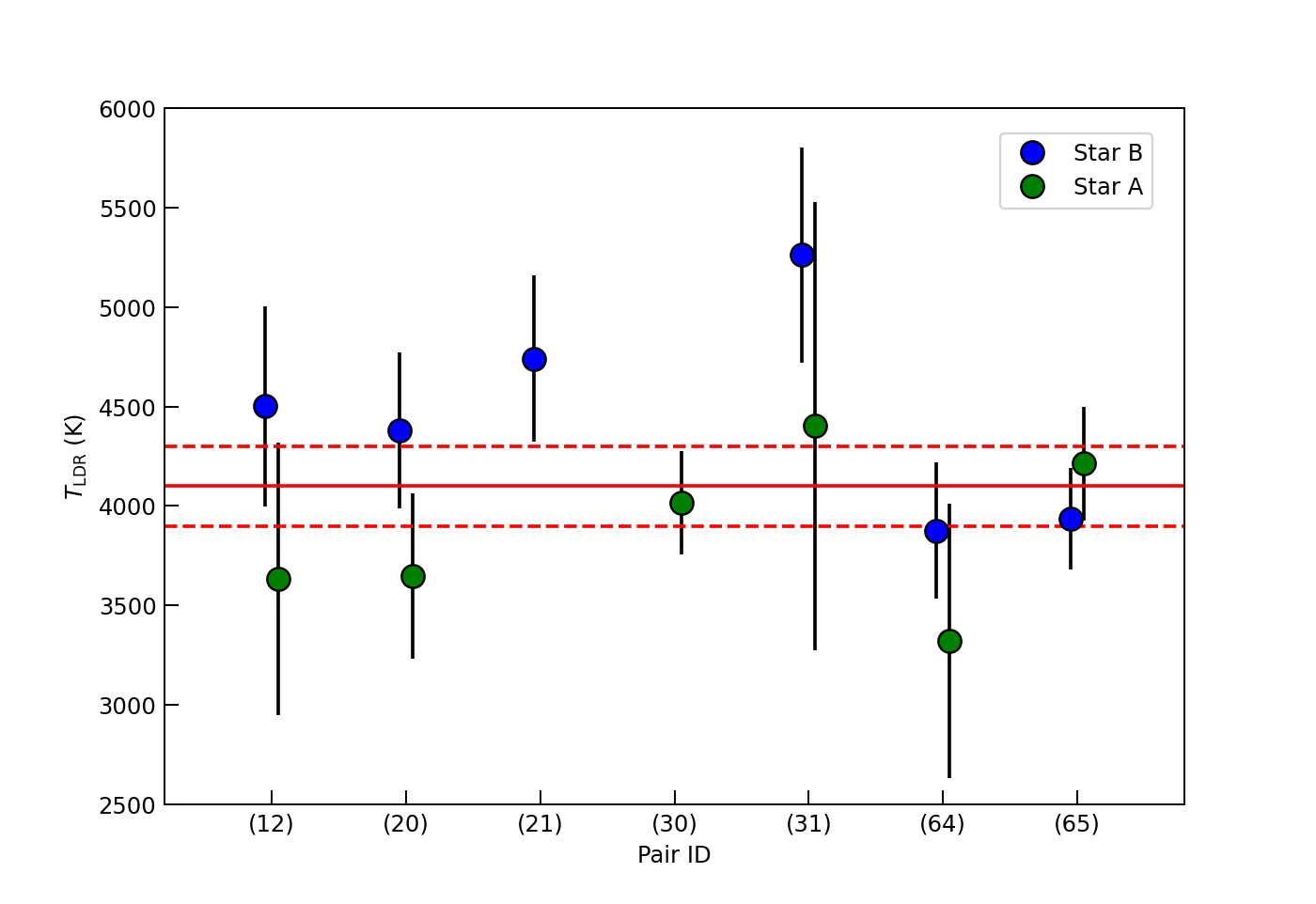}
\caption{
LDR-based temperatures estimated with line pairs taken from Taniguchi et al. (2018). The solid and dashed lines indicate the temperature and its error used for measuring chemical abundances of the two stars. 
}\label{fig:comp}
\end{figure}

\subsection{Observation}
On June 10th, 2023, we used 
the WINERED spectrograph \citep{Ikeda-2022,Matsunaga-2023}
attached to the Magellan Clay telescope in Chile to
obtain high-resolution spectra of two stars.
WINERED is a near-IR high-resolution spectrograph covering
0.90–1.35 $\mu$m ($z^{\prime}$, $Y$, and $J$ bands), with a resolution of $R = \lambda /
\Delta\lambda = 28,000$ with the WIDE mode (Ikeda et al. 2022). The raw spectral data were reduced with the WINERED
Automatic Reduction Pipeline (WARP\footnote{https://github.com/SatoshiHamano/WARP/}, version 3.8). We confirmed that the broadening width in the final spectra is as small as 12 km~s$^{-1}$, which can be explained by the combination of the instrumental resolution (10.7 km~s$^{-1}$) and a typical macroturbulence of about 5 km~s$^{-1}$. 
These stars are the brightest among a few targets that were
selected as candidate members.
The two stars are the brightest, $J \simeq 13.5$~mag, of the targets
we selected, but they are rather faint targets for high-resolution
spectroscopy in the infrared. 
Our spectra with 1200s exposures for each star are
of moderate quality. While the signal-to-noise ratios are S/N = 30--50 in
the $J$ band, they are 15--20 in the $Y$ band (shorter wavelengths) 
because of severe interstellar extinction. Therefore, our analysis relied
mainly on the $J$-band part of the spectra (Fig.~\ref{fig:spectra}).  
The spectra of the two stars show a striking resemblance to
each other (Fig.~\ref{fig:spectra}), with the same absorption lines
appearing at the same wavelengths and exhibiting very similar depths.
This already indicates that the two target stars are red giants
with very similar characteristics, belonging to
a common stellar group, VVV CL002.

\subsection{Measurements of radial velocities}

\begin{table}[htbp]
\caption{Observed target stars  and mean globular cluster parameters}\label{tab1}%
\begin{tabular}{@{}llll@{}}
\hline
Parameter & Star A  & Star B & Mean \\
\hline
ID (VIRAC2)       & \small{9327562027332}   & \small{9327562015390}   &   \\
RA (J2000)      & 17:41:06.65     & 17:41:04.94     &   \\
Dec (J2000)     & $-$28:50:23.1   & $-$28:50:44.4   &   \\
$K_{\rm s}$ (mag) & 11.25           & 11.21           &   \\
$J-$K$_{\rm s}$ (mag) & 2.20        & 2.25            &   \\
$V_{\rm helio}$ (km\,s$^{-1}$) & $-27.3\pm 0.2$         & $-27.3\pm 0.2$         & $-27.3\pm 0.1$  \\
$V_{\rm LSR}$ (km\,s$^{-1}$)   & $-17.0\pm 0.2$         & $-17.1\pm 0.2$         & $-17.1\pm 0.1$  \\
${\rm [Fe/H]}$   & $-0.68\pm 0.37$ & $-0.39\pm 0.41$ & $-0.54\pm 0.27$ \\
${\rm [Mg/H]}$   & $-0.26\pm 0.19$ & $-0.14\pm 0.12$ & $-0.19\pm 0.10$ \\
${\rm [Si/H]}$   & $-0.12\pm 0.29$ & $-0.02\pm 0.25$ & $-0.07\pm 0.19$ \\
\hline
\end{tabular}
\end{table}

We measured radial velocities using the cross correlation technique
involving the model synthetic spectra and telluric absorption spectra
\citep{Matsunaga-2015}. For this analysis, we include some $Y$-band
echelle orders together with $J$-band orders, in which telluric lines and
stellar lines are well mixed. 
We obtained almost identical velocities for the two stars (Table~\ref{tab1}),
strongly supporting the membership to VVV~CL002 combined with
the proper motions (Fig.~\ref{fig:virac}).

\subsection{Measurements of chemical abundances}

In spite of the limited S/N, we are able to measure the abundances of
Fe, Mg, and Si. First, we estimated the effective temperature employing
the line-depth ratios \citep{Taniguchi-2018}. We could measure the ratios of 6--7 line pairs taken from Taniguchi et al. (2018) as illustrated in Fig.~\ref{fig:comp}.
Adopting an error of
200~K, we use the approximate value of the temperatures,  $T_\mathrm{eff}=4100$\,K,
for both stars in the following analysis.
It is impossible to determine other stellar parameters due to the limited quality of the current spectra. Considering the similarity to Arcturus,
we used the following parameters: $\log g=1.5$, and the microturbulent velocity $\xi=1.5$\,km\,s$^{-1}$ \citep{Fukue-2021}.

Then, we determined the chemical abundances 
by fitting model spectra to individual absorption lines that were isolated
from telluric lines and other strong stellar lines. 
The measurements were successfully done for
about 10 Fe I lines, 7 Si lines, and 4 Mg lines, leading to 
the abundances listed in Table~\ref{tab1}. The errors in the [X/H] values of individual stars are given by
the combination of the standard deviations of the line-by-line abundances
and the systematic errors caused by uncertainties in stellar parameters
such as $T_\mathrm{eff}$.
The line-by-line random errors dominate the total errors in all cases, i.e., the three elements in the two stars.  Among the systematic errors, the errors from the log g uncertainty tends to be the largest error source, 0.6--0.9 dex in [X/H], but the Teff uncertainty has the highest impact in [Si/H], $\sim$0.1 dex. The trends of the systematic errors are similar to what was found for Arcturus (Fig. 7 in Fukue et al. 2021), but the errors in each stellar parameter, especially logg, are significantly larger for our targets. 
Finally, we took weighted means and their errors 
of the [X/H] values of the two stars to discuss
the chemical abundances of the cluster (Table~\ref{tab1}).
In fact, we obtained almost identical velocities 
and common chemical abundances within the errors for
the two stars (Table~\ref{tab1}).
Such close agreements are unexpected for field stars
in the bulge characterized by a large velocity dispersion 
and a wide metallicity distribution \citep{RecioBlanco-2018,Zasowski-2019,Schultheis-2020,Geisler-2021,Queiroz-2021}.
In addition, the spectra of these two stars are similar
to the high S/N spectrum of the prototype red giant
Arcturus (${\rm [Mg/H]}=-0.2$, ${\rm [Si/H]}=-0.2$, ${\rm [Fe/H]}=-0.5$ \citep{Fukue-2021}),
supporting the metallicities and the $\alpha$ enhancement of our two stars.
There are absorption lines of some other elements seen in the obtained spectra, as indicated in Fig. 1. It would be valuable to measure the abundances of such elements, but we limited ourselves to the three elements (Fe, Si, Mg) for which simple analysis with the limited-S/N spectra was sufficient. For example, the number of useful Ca lines is more limited than Mg because of blends by telluric lines and stellar lines and also because only a few lines have moderate depths appropriate with the limited S/N values. Ti tends to be affected by NLTE. Measuring C and/or N abundances with CN would require CO and OH lines together, but none of these lines are available in the WINERED range. 

\section{Cluster orbit}

\begin{figure*}[htbp]%
\centering
\includegraphics[width=\hsize]{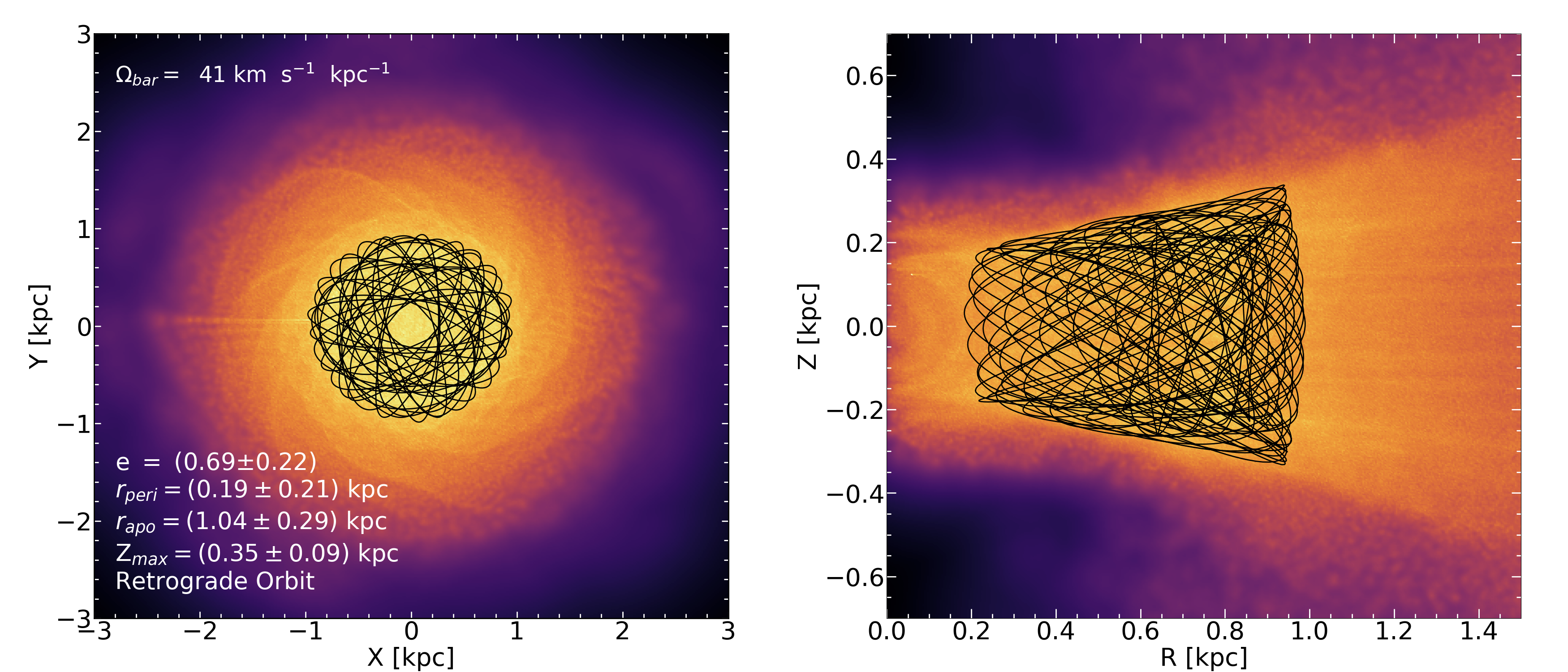}
\caption{
Orbit computed for VVV~CL002 (black line), overlaid on the probability densities of orbits projected on the Galactic plane (left) and height above the plane z versus Galactocentric radius (right). 
Lighter colours indicate more probable regions of space, that are more frequently sampled by the simulated orbits. 
}\label{fig:orbits}
\end{figure*}

We computed the orbit using the 3D barred Galaxy steady-state potential model of the GravPot16 code \citep{Fernandez-Trincado-2022}. 
We ran the orbital simulation considering different bar pattern speeds, $\Omega_{bar} = 31$, $41$, and $51$ km\,s$^{-1}$\,kpc$^{-1}$ \citep{Sanders-2019}. 
For VVV~CL002, we obtained 10,000 orbits adopting a simple Monte Carlo re-sampling, where the uncertainties 
in the input coordinates ($\alpha$, $\delta$), proper motions, radial velocities, and distance errors were randomly propagated as $1 \sigma$ variations in the Gaussian Monte Carlo re-sampling.
Fig.~\ref{fig:orbits} shows the computed orbits using $\Omega_{bar} = 41$\,km\,s$^{-1}$\,kpc$^{-1}$, displayed as probability densities of orbits projected on the equatorial Galactic plane (left panel) and height above the plane z vs Galactocentric radius  in kpc (right panel). The lighter colours indicate more probable regions of space, that are travelled more frequently by the simulated orbits. 
According to the result of this calculation,
VVV~CL002 has a retrograde orbital configuration of relatively high eccentricity ($e=0.69 \pm 0.22$), with perigalactocentric and apogalactocentric distances well inside the Galactic bulge at $R_{peri} = 0.19 \pm 0.21$~kpc and $R_{apo} = 1.04 \pm 0.29$~kpc, respectively, and with moderate vertical excursions from the Galactic plane ($|Z_\mathrm{max}| = 0.35 \pm 0.09$~kpc).
%
%
It is important to note that any adopted heliocentric distance, our simulations confirm that kinematically VVV~CL002 now belongs to the bulge, instead of being a halo globular cluster in a very eccentric orbit that is merely passing by the inner bulge.


\section{Discussion}

The radial velocity we obtained allows to give a strong constraint to
the full 3D motion of the cluster. We calculated 10,000 orbits
of VVV~CL002, adopting a Monte Carlo re-sampling
with the uncertainties in input parameters
such as distance, radial velocity and proper motion taken into account. 
We found that VVV~CL002 has a retrograde orbital configuration of
relatively high eccentricity ($e=0.69 \pm 0.22$), with perigalactocentric
and apogalactocentric distances well inside the bulge,
$R_\mathrm{peri} = 0.19$\,kpc and $R_\mathrm{apo} = 1.04$\,kpc (Fig.~\ref{fig:orbits}).
The retrograde motion is not unique, as there exist a few other retrograde globular clusters enduring the harsh density of the Galactic bulge \citep{Romero-Colmenares-2021, 
Perez-Villegas-2020, Garro-2023}. 
However, the orbit of VVV~CL002 is tighter than the orbits of all other known globular clusters \citep{Perez-Villegas-2020}. 
No globular cluster is expected to survive over its lifetime
($>$10\,Gyr) in such proximity to the Galactic centre \citep{Gnedin-2014}. 

\begin{figure*}[htbp]%
\centering
\includegraphics[width=\hsize]{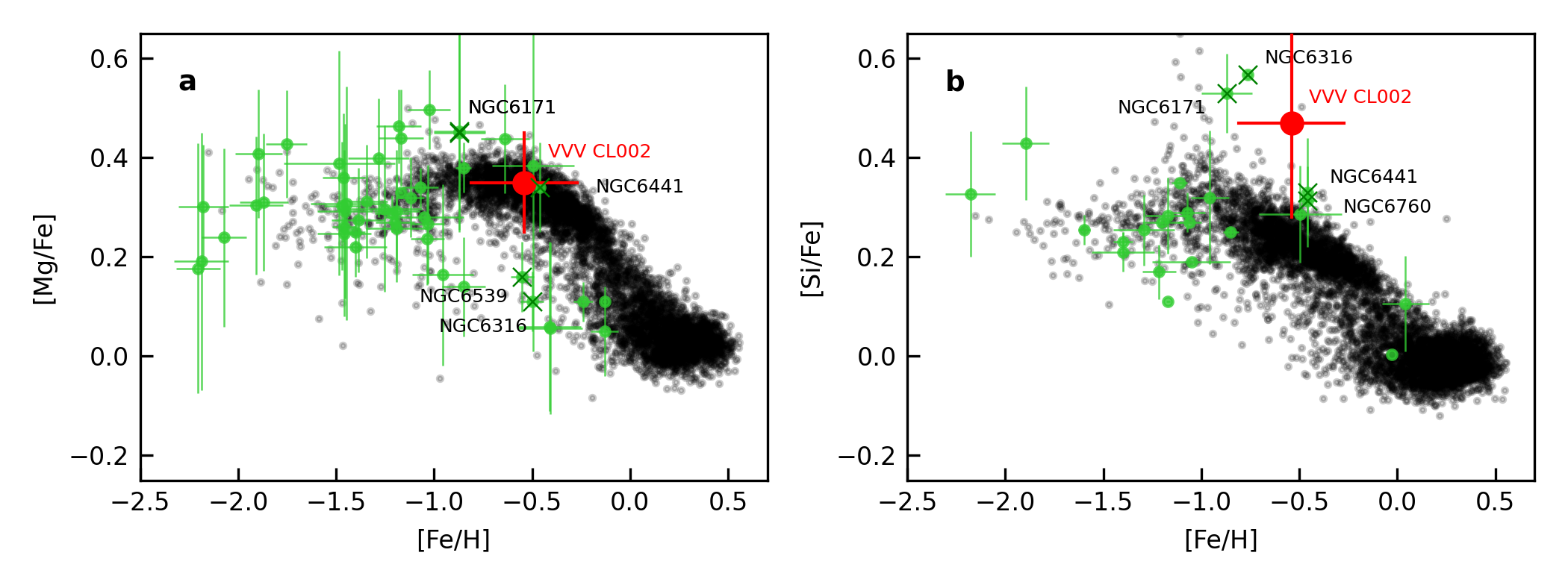}
\caption{
[Mg/Fe] versus [Fe/H] (left) and [Si/Fe] versus [Fe/H] (right) for the globular cluster VVV~CL002 (red circle) compared with the abundances of other known globular clusters indicated by green circles \citep{Meszaros-2020,Carretta-2009,Dias-2016,RecioBlanco-2018,Barbuy-2018,Barbuy-2021,Geisler-2021,Schiavon-2023}
and bulge field stars indicated by small dots \citep{Abdurrouf-2022,Queiroz-2023}
The globular cluster 
mean abundances with their respective errors are all from mid/high resolution optical and IR spectroscopy from the literature as compiled by Garro et al. (2023, submitted).
Also a few representative globular clusters that bracket the VVV-CL002 abundances are marked with crosses and labelled for comparison.
}\label{fig:abundances}
\end{figure*}

In order to unveil the mysterious history of this cluster,
its chemical abundance plays an essential role.
The abundances of the two stars agree within the errors,
giving an estimate of the cluster's metallicity,
[Fe/H]$=-0.54 \pm 0.27$, consistent with a previous photometric determination
([Fe/H]$=-0.4$, \citep{MoniBidin-2011}), and [$\alpha$/Fe] $\simeq +0.4$ (Table~\ref{tab1}).
The high $\alpha$ enhancement is connected to rapid chemical evolution
dominated by core-collapse supernovae rather than type Ia supernovae \citep{Sharma-2021}.
This confirms that VVV~CL002 is an old globular cluster 
formed together with other clusters and field stars
present today in the Galactic bulge (Fig.~\ref{fig:abundances}), rather than
 a younger open cluster or the remains from an (already-disrupted) dwarf galaxy \citep{Hughes-2020}.
Furthermore,
using a state-of-the-art technique for estimating stellar birth radii ($R_\mathrm{birth}$) within the Galaxy \citep{Minchev-2018}, 
recent studies demonstrated that stars with
relatively low metallicity (among bulge stars)
and high $\alpha$ enhancement were formed more
in the outer part, 3--6\,kpc in $R_\mathrm{birth}$
\citep{Lu-2022,Ratcliffe-2023}.
This brings us to a scenario where VVV~CL002 was formed
at a relatively large $R_\mathrm{birth}$ and started
to fall towards the centre recently. 
It is probably doomed to continue spiralling into the inner parsecs and be destroyed in the not-so-distant future.
This cluster sheds light on the intriguing survival and migration mechanisms of globular clusters,
whereas many less-characterized globular clusters and candidates 
are within a couple of kilo-parsecs from the centre.
Demand is high for near-infrared high-resolution spectroscopy
of such clusters, which has been handicapped 
due to severe interstellar extinction.

\begin{acknowledgements}
      This paper is based on the WINERED data gathered with
      the 6.5 meter Magellan Telescope located at Las Campanas Observatory, Chile.
      This research is supported by JSPS Bilateral Program Number JPJSBP120239909. 
      The observing run in 2023 June was partly supported by KAKENHI 
      (grant No 18H01248). We also acknowledge Scarlet S. Elgueta and
      Rogelio R. Albarracin for supporting the observations. 
      WINERED was developed by the University of Tokyo and the Laboratory
      of Infrared High-resolution Spectroscopy, Kyoto Sangyo
      University, under the financial support of KAKENHI
      (Nos. 16684001, 20340042, and 21840052) and the MEXT
      Supported Program for the Strategic Research Foundation at
      Private Universities (Nos. S0801061 and S1411028).
      We gratefully acknowledge the use
      of data from the ESO Public Survey program IDs 179.B-2002 and
      198.B-2004 taken with the VISTA telescope and data products
      from the Cambridge Astronomical Survey Unit.
      D.M. acknowledges support by the ANID BASAL projects ACE210002 and
      FB210003, by Fondecyt Project No. 1220724, and by CNPq/Brazil
      through project 350104/2022-0. 
      J.G.F.-T. acknowledges support provided by Agencia Nacional de Investigaci\'on y Desarrollo de Chile (ANID) under the Proyecto Fondecyt
      Iniciaci\'on 2022 Agreement No. 11220340, and from the Joint Committee ESO-Government of Chile 2021 under the Agreement No. ORP
      023/2021, and from Becas Santander Movilidad Internacional Profesores 2022, Banco Santander Chile.
      R.K.S. acknowledges support from CNPq/Brazil through projects 308298/2022-5 and 350104/2022-0.
      E.R.G. acknowledges support from ANID PhD scholarship No. 21210330.
\end{acknowledgements}

%
%

\bibliographystyle{aa} 
\bibliography{biblio} 

%
%
%
%
%
%
%
%
%

\end{document}